\documentclass{JHEP3}
\usepackage{epsfig}
\usepackage{graphicx}
\usepackage{dcolumn}
\usepackage{bm}
\newcommand{\be}{\begin{equation}}
\newcommand{\ee}{\end{equation}}
\def\bea{\begin{eqnarray}}
\def\eea{\end{eqnarray}}

 \def\be{\begin{equation}}
\def\ee{\end{equation}}
\def\bea{\begin{eqnarray}}
\def\eea{\end{eqnarray}}

\def\lesssim{\mathrel{\hbox{\rlap{\hbox{\lower4pt\hbox{$\sim$}}}\hbox{$<$}}}}
\def\gtrsim{\mathrel{\hbox{\rlap{\hbox{\lower4pt\hbox{$\sim$}}}\hbox{$>$}}}}

\title{N=2 Supersymmetry and de Sitter space}

\author{Renata Kallosh \\
    Department of Physics, Stanford University, Stanford, CA 94305,
USA\\
e-mail: kallosh@stanford.edu}
\received{\today}       
 \preprint{SU-ITP-01/37\\  ~hep-th/0109168\\ \today}

\abstract{We present a (unique?) possibility of de Sitter solution in the framework of N=2 supersymmetry (hypersymmetry). We show that a model with a vector and a charged hypermultiplet has a hybrid inflation type potential. It leads to a slow-roll regime in de Sitter type background with all supersymmetries broken spontaneously. Beyond the bifurcation point the waterfall stage abruptly brings the system into a ground state with 2 unbroken supersymmetries. De Sitter stage exists under condition that the superconformal $SU(2,2|2)$ symmetry of the model is softly broken down to N=2 supersymmetry by the vev of the Killing prepotential triplet $P^r$.  This hybrid hypersymmetry model may describe P-term inflation and/or acceleration of the universe at the present epoch.
}

\begin{document}

\section{Introduction}

Current cosmological observations indicate that there could be a stage of a rapid  accelerated expansion (inflation) in the very early universe  \cite{inflation}. Such a stage could start immediately after the big bang and it could last as long as $10^{-35}$ seconds (in the simplest inflationary models). Moreover, observations indicate that few billion years after the big bang (half-way to the present epoch) the universe entered a second stage of accelerated expansion. The rate of acceleration (the Hubble constant) now is approximately 60 orders of magnitude smaller than during the stage of inflation in the early universe. For a discussion of the observations and their theoretical interpretation see e.g. \cite{CMB}.

Usually it is assumed that the energy-momentum tensor during inflation is dominated by potential energy density of a scalar field,  $T_{\mu\nu} \approx g_{\mu\nu} V(\phi)$ with $V >0$. This condition requires that the scalar field moves very slowly, $\dot\phi^2/2 \ll V(\phi)$ \cite{book}. The limiting case $\dot\phi = 0$ corresponds to de Sitter space with a positive cosmological constant. Similarly,  current cosmological acceleration can be explained either by a positive vacuum energy $V$ (cosmological constant)  or some kind of slowly rolling scalar field in a near de Sitter background with $\dot\phi^2/2 \ll V(\phi)$  \cite{quint}. 

For the sake of the argument, let us assume that inflation and  acceleration of the universe will be supported by future experiments and moreover,  the supersymmetric particles will be  detected and supersymmetry confirmed. What kind of theories can be used to explain such experiments? 

The relation between supersymmetry, supergravity and positive cosmological constant was considered as problematic for a long time \cite{Ferrara:1977sc}-\cite{Pilch:1985aw}. It has been recognized long ago that  in extended 
supersymmetric theories de Sitter solutions are unstable solutions with broken supersymmetry\cite{Gates:1983ct,Hull:1985ea}.

Recently it was pointed out that it may not be easy to describe de Sitter space and accelerating universe in M/string theory  \cite{Banks:2000fe}-\cite{Fischler:2001yj}. 
It was particularly stressed by Witten in \cite{Witten:2001kn} that no clear-cut way to get de Sitter space from M/string theory is visible. Very  recent discussion of these issues can be found in \cite{Spradlin:2001pw,Hull:2001ii,Townsend:2001ea}.

The new angle from which we would like to look at the old issue is: {\it how unstable is de Sitter space in supersymmetric theories? When can this instability  be used in the cosmological context?} There are few instances in N=8 and N=4  gauged supergravities related to M/string theory, where de Sitter classical solutions have been found in  the past. We shall discuss them in a separate publication \cite{Kallosh:2001gr}. 

Theories of N=1 supergravity coupled with N=1 matter have an opposite problem. There is an unlimited number of possibilities to find  stable de Sitter solutions. The choice of the K\"{a}hler potential $K(z, \bar z)$, the holomorphic superpotential $W(z)$ and D-terms allows  various type of potentials with de Sitter vacua. There is no clear preference for one or another choice.

N=2 supersymmetry and N=2 supergravity have not attracted much attention in this respect, to the best of our 
knowledge \footnote{ A model of inflation based on the soft-breaking of
N=2 supersymmetry was studied in  \cite{Garcia-Bellido:1998mq}.}. The purpose of this note is to show that a very interesting and rather unique possibility opens up here. We will start with a global $SU(2,2|2)$-superconformal gauge theory which hopefully may be related, via Maldacena conjecture, to string theory compactified on $ADS_5\times M^5$-space, where $M^5$ is  some supersymmetry breaking manifold. The model we will study is a well known N=2 QED with one vector multiplet and one charged hypermultiplet. Presumably, it is a part of a more general theory, but the main new features of the  cosmological scenario develop in this sector. We add the FI term $\xi$ equal to $ \sqrt {2\Lambda} $ where $\Lambda$ is a desirable value of the cosmological constant. This will explicitly break the superconformal symmetry but N=2 supersymmetry will be intact. The theory can be easily coupled consistently to N=1 supergravity; the coupling to N=2 supergravity may face some topological obstructions for non-vanishing constant $\xi$.

We find that the potential of our N=2 model  is  of the type used in  hybrid inflation \cite{Linde:1991km} with de Sitter stage and a waterfall stage. The analogous hybrid potentials were used in  F-term inflation \cite{Copeland:1994vg,Dvali:1994ms} and D-term inflation \cite{Stewart:1995ts,Binetruy:1996xj,Halyo:1996pp},  see \cite{Lyth:1999xn} for a review. Even though the N=2 theory   differs substantially from the one used in  F-term inflation, in  the simplest case of one hypermultiplet  our inflationary potential at the classical level exactly  coincides with the F-term potential of \cite{Dvali:1994ms}. It also coincides with  the D-term potential  of N=1 theory \cite{Binetruy:1996xj}
under condition that the gauge coupling $g$ is related to the Yukawa coupling $\lambda$ as follows: 
\be
\lambda^2=  2 g^2 \ . \nonumber
\ee
The potential depends on vev's of at least 3 scalar fields (or more), one scalar from the vector multiplet and two from the hypermultiplet.

The mechanism of hybrid inflation may apply not only to early universe and inflation era, but also to the present stage of the accelerating universe (for a different set of fields and parameters).
For the early universe inflation in this scenario $\sqrt {\xi\over g} \sim 10^{-2} M_P$ with the relevant scale $10^{-30}$ cm  but for the present cosmological constant
$\sqrt {\xi} \sim 10^{-30} M_P$ with the length scale of the order of $10^{-2}$ cm. There will be a long slow change of an `inflaton field' in the near de Sitter space (the space with $T_{\mu\nu} \approx g_{\mu\nu} V$)
 with  broken susy. At the critical point where the vev of the inflaton will reach the value $\sqrt{ \xi\over g}$, the potential will switch into the waterfall stage  and very fast the ground state with 2 unbroken supersymmetries and vanishing cosmological constant will be reached.

The fundamental role in this cosmological scenario is played by the non-vanishing vev of the third component  of the triplet of the Killing prepotentials $P^r$ of N=2 theory with special and hyper-K\"{a}hler geometry.
(In a simplest case this triplet is an auxiliary field of the N=2 vector multiplet). 
Since inflation and acceleration of the universe in this model appear due to the $P$-term in N=2 supersymmetry, we will call it $P$-term inflation.

This model have several different advantages over the models based on N=1 supersymmetry. First of all, it is much more restrictive than in the more familiar cases of F-term  and/or D-term inflation, where one can choose many different superpotentials without a clear guiding principle. In the N=2 theory one does not have this freedom. As a reward, one can show that all non-gravitational quantum corrections in this theory starting from the second loop are finite \cite{Grisaru:1982zh}.

The basis of this model is the
$SU(2,2|2)$-superconformal symmetry of the gauge theory which may live at the  boundary of the $ADS_5$ space related to string theory. In this way one may try to connect both inflation and accelerating universe to string theory. 

The slow de Sitter type evolution in this model is not eternal. After passing the critical point the Minkowski ground state is naturally reached where one of the hypermultiplet scalars acquires a vev $ \sqrt {2\xi\over g}$ and the unbroken N=2 supersymmetry is restored\footnote{A non-supersymmetric model of hybrid quintessence  where the acceleration is not eternal has been suggested recently by Halyo \cite{Halyo:2001fb}.}. 
At this ground state a vector multiplet `eats' a hypermultiplet and transforms into one massive N=2 vector multiplet, which is a super-Higgs effect in N=2 theories. All fields have the same   mass $m^2=2 g\xi$. This overcomes the problem of eternal quintessence pointed out for one scalar field in supersymmetric theories  in \cite{Hellerman:2001yi} since in our case the system does relax into a zero-energy supersymmetric vacuum.

\section{De Sitter solutions in N=1 supergravity}

Before discussion of coupling to supergravity of our N=2  model, let us make a short overview of de Sitter solutions in N=1 supergravity. 
We will point out necessary and sufficient conditions for N=1 supergravity to have a de Sitter solution.

The well known potential depends on K\"{a}hler potential $K(z, \bar z)$, on holomorphic superpotential $W(z)$ and on a holomorphic function $f_{\alpha \beta}(z)$ in kinetic terms for vector multiplets   \cite{Cremmer:1983en}: 
\begin{eqnarray}
V=V_F+V_D= e^{ K\over M_P^{2}}\left[ (
{\cal D}^iW)K^{-1}{}_i{}^j({\cal D}_jW^*) -3{WW^*\over M_P^{2}})\right]
+ {1\over 2} (\rm Re\,  f_{\alpha \beta}) D^\alpha D^\beta \ .
\label{Vtotal}
\end{eqnarray}
The potential is not positive definite, the negative contribution is proportional to the square of the gravitino mass,
$$ M_{\rm gravitino} =   {W\over M^2_P} e^{ K\over 2 M_P^{2}}\ .$$ 
To find a de Sitter solution one has to find some constant values of all scalars at which the potential has an extremum with 
\be
3|W|^2 < |F|^2 + {1\over 2} |D |^2 e^{-{ K\over M_P^{2}}} \ .
\ee
where $|F|^2 \equiv (
{\cal D}^iW) K^{-1}{}_i{}^j({\cal D}_jW^*)$.
Clearly, any solution with vanishing or very small superpotential $W$ at the critical point  and  with non-vanishing F-term, 
 derivative of the superpotential, and/or non-vanishing D-term will give a de Sitter solution. There are many ways to get it. 

Consider an example:
\be
K= z z^* +y y^* \ , W= (z^2 +a) y \ .
\ee
The critical point of the potential at $z=y=0$ gives a de Sitter solution. The value of the potential at this critical point is
\be
V_{0}= a^2 \ .
\ee
There are many known examples with vanishing chiral multiplets contribution, in all such cases D-terms will give a positive cosmological constant. Thus N=1 supergravity needs some motivation from any version  of a more fundamental theory like  M/string theory about the choice of ${\cal G}$ and relevant physics.

\section{Hybrid hypersymmetry  model}

To reveal the essential features of hypersymmetry\footnote{`When a  French super-march\'{e} carries not only food and drink but also car spares, garden furniture and ladies' underwear, it becomes an hyper-marche\'{e}. Correspondingly, P. Fayet  called N=2 supersymmetry hypersymmetry', as explained in  \cite{Sohnius:1985qm}.} in cosmology we will use a simple N=2 model  suggested by Salam and Strathdee \cite{Salam:1975wa} and Fayet \cite{Fayet:1976yi} (SSF model).  The reader familiar with Witten-Seiberg paper \cite{Seiberg:1994aj} may find it easy to recall the model  as a warm-up example there: QED with matter, i.e.
an abelian gauge theory with N=2 supersymmetry and charged matter hypermultiplets.  We will first study the version of the theory with massless hypermultiplets and explain the massive case later. In what follows we consider one hypermultiplet, more general case will have analogous important features and,  perhaps,  more. The supersymmetric ground state of this model is well known, the existence of the local  valley with positive potential was not  noticed before in this N=2 model.

\subsection{$SU(2,2|2)$ superconformal part of the model}
In terms of N=1 superfields the superconformal part of the model is rather simple, in notation of  \cite{Fayet:1976yi} it is: 
\be
{\cal L}= {\cal L}_0 + [S^* e^{2gV}S + T^*e^{-2gV} T + N^* N]_D + [4g T^*S N]_F \ .
\ee 
Here  ${\cal L}_0$ is the kinetic term for the $U(1)$ gauge multiplet, $N$ is the neutral chiral supermultiplet whereas $S$ and $T$ are charged chiral superfields of the opposite charge. There is only one coupling constant, both for gauge coupling and for Yukawa coupling.

It will be  useful for our purpose to present this theory in terms of component fields with off-shell rigid  N=2 supersymmetry, following  \cite{Sohnius:1985qm,West:1990tg}. 

We have an abelian N=2 {\it   gauge multiplet} which belongs to a representations of rigid $SU(2)$, the antisymmetric tensor $ \varepsilon_{AB} $, $A,B=1,2$,  is used to raise and lower the $SU(2)$ indices. The multiplet consists of two scalars $A$ and $B$, a vector $A_\mu$, (all singlets in $SU(2)$),  a spin-1/2 doublet $ \lambda^{ A}= \varepsilon^{AB} \gamma_5 C \bar \lambda_B^T$ (gaugino) and an  auxiliary field
$P^{r}$, triplet in $SU(2)$. This simple model is a particular case of  the general setting of N=2 gauge theories and gauged supergravity where the triplet of Killing prepotentials $P^r $ plays an important role in special geometry and in  hyper-K\"{a}hler (quaternionic for supergravity) manifolds. 

A  hypermultiplet, N=2 {\it   matter multiplet} has two complex scalar fields forming a doublet under $SU(2)$, $\Phi^A$ and $\Phi_A= (\Phi^A)^*$   and a spin 1/2 field $\psi$, singlet under $SU(2)$ (hyperino). There is also a doublet of dimension 2 auxiliary fields, $F^A$ with $F_A= (F^A)^*$ .
The superconformal action is
\bea
\label{model}
{\cal L}_{s.c.}&=& {1\over 2} [(\partial_\mu A)^2+ (\partial_\mu B)^2]    +{i\over2}\,  \bar \lambda_A  \not\!{\partial} \lambda^A -{1\over 4}F_{\mu\nu}^2 + {1\over 2} \vec P^2 \nonumber\\
\nonumber\\
&+& {1\over 2} D_\mu \Phi^A D^\mu \Phi_A +i\, \bar \psi \not\!{D } \psi +  F^A F_A  \\
\nonumber \\
&+& i g \Phi^A \bar \lambda_A \psi   -i g \bar \psi  \lambda^A \Phi_A  - g \bar \psi ( A - \gamma_5 B) \psi + {g\over 2}   \Phi^A \vec \sigma _A{}^B \vec P \Phi_B - {g^2\over 2} \Phi^A  (A^2 + B^2)\Phi_A \ . \nonumber 
\eea
The covariant derivatives on the hypers are
\bea
D_\mu \Phi_A &=& \partial_\mu \Phi_A  +i g A_\mu \Phi_A \ , \nonumber\\
D_\mu \Phi^A&=& \partial_\mu \Phi^A -i g A_\mu \Phi^A \ , \nonumber\\
D_\mu \psi&=& \partial_\mu \psi +i g A_\mu \psi \ .
\label{cov}\eea
The first line in (\ref{model}) is for the  vector multiplet, the second one is for the hypermultiplet. Note that all 6 scalars, $ f \equiv\{ A,B, a_1, b_1, a_2, b_2\}$, where $\Phi_1=a_1+ i b_1$ and $\Phi_2= a_2+ i b_2$ have canonical kinetic terms of the form
${1/2} (\partial f)^2$.
The third line  includes terms which must be added to the action simultaneously with covariantization of derivatives. All term  depending on $g$ describe the gauging. When the gauge coupling $g$ is vanishing, the theory is a non-interacting theory of a vector multiplet and a neutral hypermultiplet.

The model has 2 rigid  supersymmetries (hypersymmetry) with the  Majorana spinors $\varepsilon^A= \varepsilon_{AB} \gamma_5 C  \bar \varepsilon_B^T$. 
The fields of the gauge multiplet transform as follows:
\begin{eqnarray} 
\label{eq:gauge2trans}
\delta A_{\mu} &  = & 
 i\,\bar{\epsilon}_A\,\gamma_{\mu}\,\lambda^A \,, \qquad \delta A      = 
 i\,\bar{\epsilon}_A\,\lambda^A \,,\qquad \delta B      = 
  i\,\bar{\epsilon}_A\,\gamma^5\,\lambda^A\,, 
 \nonumber \\
\delta \lambda^A &  = & 
  -\frac{i}{2}\,\sigma^{\mu \nu}\,\epsilon^{ A} F_{\mu \nu} -  
  \gamma^{\mu}\,\partial_{\mu}\,(A + 
  \gamma_5\,B)\,\epsilon^{A} - 
  i\,\epsilon^{B}\,\vec{\sigma}_{B}{}^{A}\,\vec{P}\,, 
\label{gaugino} \\
\delta \vec{P} &  = & 
  \epsilon_B\,\vec{\sigma}_{A}{}^{B}\,\gamma^{\mu}\,\partial_{\mu} \lambda^A\ . \nonumber
\end{eqnarray}
The supersymmetry transformations of the fields of the
 hypermultiplets are
\begin{eqnarray}
\delta \Phi_{A}& = & 
  2\, \bar{\epsilon}_A\, \psi \,,\qquad \delta F_{A} =
   2\, \bar{\epsilon}_A\, (\gamma^{\mu}\,D_{\mu}+g(i\,A-i\,\gamma_5\,B)) \psi -2 g \bar{\epsilon}_B\,\lambda^B\,\Phi_{A}, \nonumber \\
\delta \psi &  = & 
  -i\, \epsilon^{\,A}\, F_{A} - i\,\gamma^{\mu}\, D_{\mu} \epsilon^{\,A}\, \Phi_{A} 
  + g(A + \gamma_5\, B)\, \epsilon^{\,A}\, \Phi_{A}\ .  
\label{hyperino}\end{eqnarray}
Using equations of motion for auxiliary fields, 
\be
 P^r=- {g\over 2} \Phi^A (\sigma^r)_A{}^B \Phi_B \ , \qquad  F^A=0 
\ee
we find the potential
\be
V_{s.c.}= {g^2\over 2} [ \Phi^\dagger \Phi (A^2 + B^2) +{1\over 4} (\Phi^\dagger \vec \sigma \Phi)^2]\ ,\nonumber
\ee
where  
$\Phi^\dagger \Phi\equiv \Phi^A \Phi_A=|\Phi_1|^2 + |\Phi_2|^2\geq 0$.
It is also possible to rewrite the potential as follows:
\be
V_{s.c.}= {g^2\over 2}[\Phi^\dagger \Phi (A^2 + B^2) +{1\over 4} (\Phi^\dagger \Phi)^2]\ .
\label{scpot}\ee
The theory has a dilatation symmetry and $U(2)$ symmetry and    it is symmetric   under the  superconformal symmetry
$SU(2,2|2)$, see \cite{deWit:1980gt,West:1990tg}. This  adds special conformal and special supersymmetry. The bosonic symmetry $SU(2,2)$ is an isometry of the $adS_5$ space which may connect this gauge theory to string theory on the boundary of the $adS_5$ space. The general
conditions for conformal invariant hypermultiplets theories have been
found in \cite{deWit:1998zg}.

\subsection{N=2 supersymmetry after soft breaking of superconformal symmetry}
There are two possibilities to break $SU(2,2|2)$ symmetry down to N=2 supersymmetry. The first one is to make a hypermultiplet massive. This is equivalent to a shift of the vev of the $A$ field, a scalar from the vector multiplet and it does not make important changes. A dramatic change in the behaviour of the potential takes place if the N=2 FI terms $\xi^r$ are added \footnote{We use the standard definition of the FI terms where the potential at the vacuum equals to  $V=  \xi^2/2$ and the $U(1)$ gauge coupling has a standard definition as shown in eq. (\ref{cov}). In application to cosmology  $\xi$ and $g$ were mixed and rescaled, see e.g. \cite{Lyth:1999xn}.}
to the theory (for abelian multiplet only). The N=2 supersymmetry of the action remains intact. The new action is:
\be
{\cal L}_{N=2}=
{\cal L}_{s.c.}+  P^r \xi^r \ .
\ee
This will change the field equation for $P^r$, which will become
\be
P^r= -{g\over 2} (\Phi^\dagger \sigma^r \Phi) - \xi^r\ .
\ee
The potential of our  hypersymmetric model becomes equal to
\be
V= g^2 [\Phi^\dagger \Phi (A^2 + B^2) +{1\over 4} (\Phi^\dagger \vec \sigma \Phi +{2\over g}\vec \xi )^2]\ .
\label{compact}\ee
We chose to have a FI term only in the direction $3$ and to have it positive
\be
\xi^r= (0,0,\xi)\ ,\qquad \xi>0 \ .
\ee
We can rewrite it in the form where it is clear  that this is a hybrid-type potential \cite{Linde:1991km}-\cite{Lyth:1999xn}:
\be
V_{N=2}= {g^2\over 2} [(|\Phi_1|^2 + |\Phi_2|^2) |\Phi_3|^2 + |\Phi_1|^2 |\Phi_2|^2+ {1\over 4}(|\Phi_1|^2 -|\Phi_2|^2+ {2\over g}\xi )^2]\ ,
\label{pot}\ee
where we introduced the notation $\Phi_3\equiv  A+iB$.
At  $\xi=0$ the potential is equal to the one of the superconformal theory, as given in (\ref{scpot}).
The potential of this N=2 supersymmetric theory coincides with the particular case  studied before in the cosmological context of the D-term inflation in \cite{Binetruy:1996xj}. There  N=1 global susy theory is defined by a superpotential $W=\lambda X \phi_+ \phi_-$ and 
the chiral multiplets $\phi_+$ and $\phi_-$ have positive and negative charges under $U(1)$. 
This potential  leads to hybrid inflation  \cite{Linde:1991km}. In the previous studies N=1 supersymmetric gauge theory was used, therefore the gauge coupling $g$ and the Yukawa coupling $\lambda$ were not related. For the special case that $\lambda=\sqrt 2 g$ the theory has N=2 supersymmetry \cite{Salam:1975wa,Fayet:1976yi}. Our notations may be compared with notation in Dvali-Binetruy \cite{Binetruy:1996xj} as follows:
$\Phi_1 =\phi_+$ and 
 $\Phi_2= \phi_-^*$ and $\Phi_3=X$, $2\xi/g=  \xi_{\rm DB} $ and $  g = 2 g_{\rm DB}$.
The potential of our N=2 supersymmetric theory coincides also with the F-term inflation N=1 theory, proposed in \cite{Dvali:1994ms}. 

\subsection{Hybrid hypersymmetry model as an example of generic N=2 gauge theory}

It may be quite important to show how this model fits  in the general class of rigid N=2 gauge theories \cite{Andrianopoli:1996vr}. Particularly since it will help to understand a coupling to supergravity. In generic case one defines the rigid special K\"{a}hler and hyper-K\"{a}hler manifolds and identifies their isometries. The gauging involves covariant derivatives on coordinates of the special K\"{a}hler manifold $z_i, \bar z^i$ and on coordinates of the  hyper-K\"{a}hler manifold $q^u$:
\bea
D_\mu z_i &=& \partial_\mu z_i + g A_\mu^I k_{Ii}\ , \nonumber\\
D_\mu \bar z^i &=& \partial_\mu \bar z^i + g A_\mu^I k_{I}^{i*}\ , \nonumber\\
D_\mu q^u &=& \partial_\mu q^u + g A_\mu k_I^u \ .
\eea
Here $k_{I}^l$ are the Killing vectors of rigid special geometry and $k_I^u$ are Killing vectors of the hyper-K\"{a}hler geometry, for each gauge group $I=1,2,\dots, n$. There are $n$ vector multiplets, so that $I=1, \dots , n$.
The special K\"{a}hler manifold of the rigid type is defined by a holomorphic section $\Omega=(Y^I,F_I)$ of a flat
bundle, where $Y^I(z)$ are special complex coordinates and $F_I$ are the
derivatives of the prepotential $F(Y)$. A K\"{a}hler potential is given by ${\cal K}= i (\bar Y^I F_I - \bar F_J Y^J)$. A derivative of the holomorphic section $\Omega$ includes the functions $f^{Ii}$ and $h_I^i$) which serve to define the period matrix.
In presence of hypermultiplets a particular N=2 gauge theory is defined in terms of the triplet of the prepotentials $P^r$, $r=1,2,3$. There are various relations between Killing vectors and prepotentials.
The gauging leads to most general  potentials of N=2 supersymmetric gauge  theory:
\be
V_{\rm susy}^{N=2}(z, \bar z, q) = g^2 (g^i{}_j k_{Ii} k^{*j}_J +4 h_{uv}k^u_I k^v_J){\bar Y}^I Y^J + g^i{}_j f^{I}_i  f^{*j J} \sum _{r=1}^{r=3} P_I^r P_J^r \ .
\label{N2rigid}\ee
Here $g^i{}_j$ is the metric for scalars of the special geometry, and $h_{uv}$ is the metric for the hypers. The potential is non-negative definite.

Our model with one gauge group, $I=1$,  gives a particular case of this potential: we have flat metrics $g^i{}_j$ and $h_{uv}$ and no Killing vectors on special manifolds, so the first term in (\ref{N2rigid}) is absent in our model. The second term in (\ref{N2rigid}) can be easily identified with the first term in our model potential (\ref{pot}) since the Killing vectors are proportional to hypers and $Y, \bar Y$ are the scalars from a vector multiplet. The second term in (\ref{pot}) is proportional to $(P^1)^2 + (P^2)^2$ and the third term in (\ref{pot}) is proportional to $(P^3)^2$. Together they reproduce the last term in eq. (\ref{N2rigid}). In a more compact form of our potential in eq. (\ref{compact}) we may identify the last terms with the last term in generic N=2 potential in (\ref{N2rigid}).

\subsection{Vacua of the hybrid hypersymmetry model}

For the vacuum solutions we chose the vanishing vector field and constant values of 3 complex scalar fields $\Phi_1, \Phi_2, \Phi_3 $. The potential (\ref{pot}) has two minima.

1. {\it Non-supersymmetric local minimum with flat directions (de Sitter valley)} at some  undefined but restricted constant value of the scalar $\Phi_3$, all other scalars vanish.
\be
\Phi_1=  \Phi_2=0 \ ,  \qquad  |\Phi_3|= |\Phi_3|_0 \ , \qquad P^3= - \xi  \ , \qquad V_0= {1\over 2} \xi^2 \ .
\ee
This is a local minimum of the potential for $ |\Phi_3|_0^2 > {\xi\over g} $.  The curvature of the potential is positive when this restriction is applied
\be
M_1^2=  g^2   |\Phi_3|_0^2 + g \xi \ , \qquad M_2^2=  g^2  |\Phi_3|_0^2 - g\xi\ , 
\ee
and there are  flat directions like $V_{,3 3 }=0$.
The supersymmetry variation of fermions (\ref{gaugino}), (\ref{hyperino}) at this vacuum is:
\bea
\delta \lambda^1 =  - i  \xi    \varepsilon^1 \ ,\qquad  \delta \lambda^2 =  + i  \xi     \varepsilon^2 \ , \qquad  \delta \psi =0  \ .
\eea
Therefore, as long as $\xi\neq 0$, all supersymmetries are broken. When coupled to gravity, this vacuum corresponds to  a {\it de Sitter solution}.

2. {\it Supersymmetric global minimum}
\be
(\Phi_3)_{ \rm susy}=(\Phi_1)_{ \rm susy}= P^3_{\rm susy}= 0 \ ,  \qquad  |\Phi_2|_{\rm susy}=\sqrt {2\xi\over g} \ , \qquad V_{\rm susy}= 0\ .
\ee

Short inspection of the supersymmetry transformation of fermions  (\ref{gaugino}), (\ref{hyperino}) shows that they all vanish:
\be
\delta \lambda_i =  0 \ , \qquad 
\delta \psi =0 \ .
\ee
Thus the absolute minimum ground state has both N=2 supersymmetries unbroken and vanishing potential. The full hypersymmetry is unbroken but the $U(1)$ gauge symmetry is spontaneously broken. The massless vector multiplet with one vector field, one complex scalar and one Dirac spinor eats the hypermultiplet with 2 complex scalars and one Dirac spinor. As a result, there is one massive vector multiplet with one massive vector field, two Dirac spinors, one complex scalars and 3 real scalars. All fields have a mass $m^2= 2 g\xi$. This is known in the literature as a hypersymmetric Higgs mechanism. When coupled to gravity, this vacuum corresponds to a solution with {\it zero cosmological constant}.

The best way to understand these two critical points of the potential is to look at the 3D plots of the hybrid potential \cite{Linde:1991km}. One of the scalars, $\Phi_1$ we will keep
unchanging at it zero value at all times. These leaves us with two scalars, the one from the vector multiplet  $\Phi_3=y$ and the one from the hypers, $\Phi_2=x$ and we take $2\xi/g=1$ for the plots.

\begin{figure}[h!]
\centering \epsfysize=8cm
\includegraphics[scale=0.9]{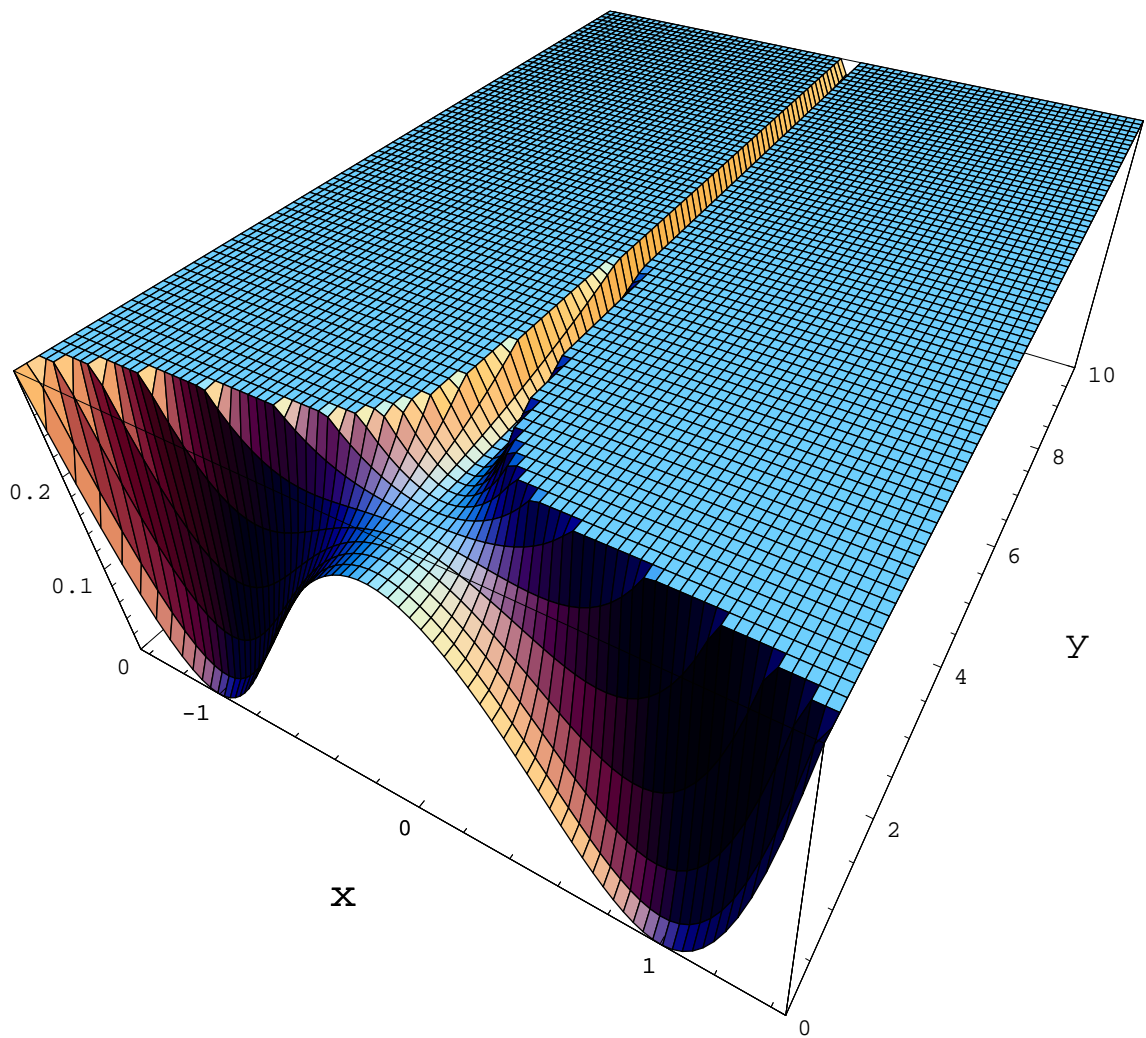}
\caption{The potential of the hybrid hypersymmetry  model. After a long de Sitter stage a local valley at $x=0$,~ $y\geq y_b$  is replaced by a waterfall stage at $x=0$, ~$0\leq  y_b$\ .} \label{onefielddistrib}
\end{figure}

\

In Fig. 1 we show the potential at $0 \leq y \leq 10 $. At large values of $y$ it has a local minimum  at $x=0$ with a flat direction along $y$ at a non-vanishing value of the potential (de Sitter valley). When coupled to gravity there will be a  long lasting de Sitter stage.  When $y$ reaches the bifurcation point $y_b=1/\sqrt 2$ the curvature of the potential at $x=0$ becomes flat and beyond the bifurcation point for $y<y_b$ one can see that at $x=0$ the potential has a maximum. This is the waterfall stage of the hybrid inflation. The field $x$ from the value $x=0$ immediately falls down to one of the positions of the absolute minima at $y=0$ and $|x|^2=2\xi/g=1$. This is a supersymmetric ground state with 2 supersymmetries unbroken.

\begin{figure}[h!]
\centering \epsfysize=8cm
\includegraphics[scale=0.9]{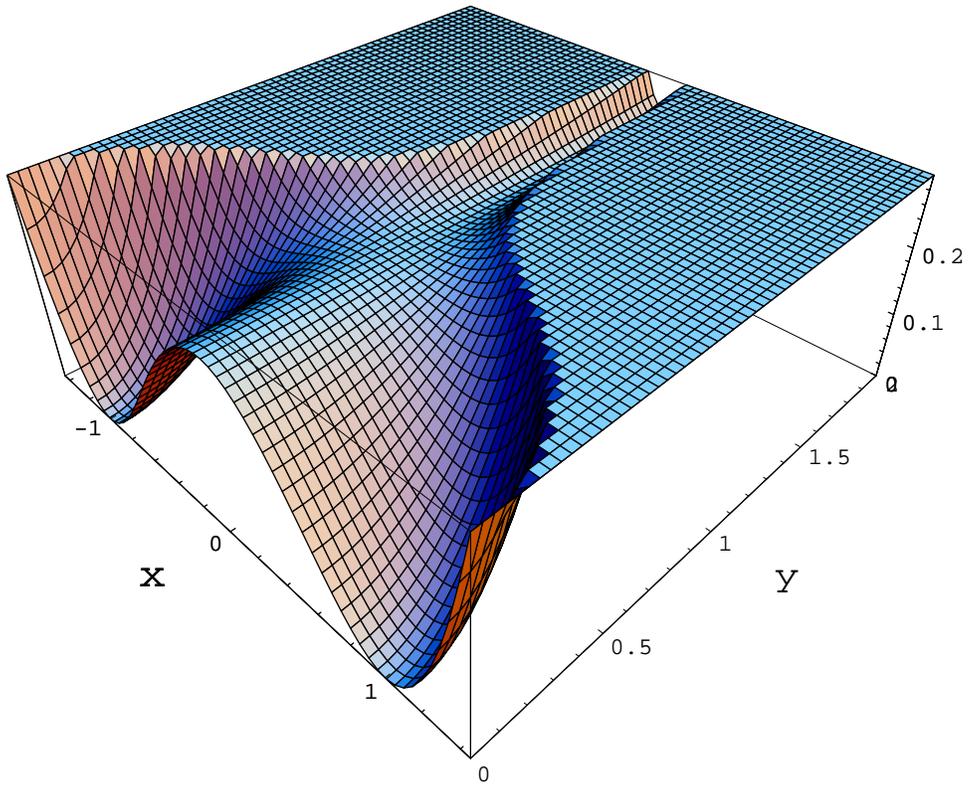}
\caption{ The potential close to the bifurcation point $x=0$,~   $y=y_b= 1/\sqrt 2$.} \label{onefielddistrib2}
\end{figure}

\
In Fig. 2  we show the potential at $0 \leq y \leq 2 $  which details the approach to the bifurcation point: the local minimum becomes a maximum after passing $y_b$. In Fig. 3 we have potential at $0 \leq y \leq y_b $ where one can see the transition of the flat potential into a maximum. In Fig. 4 we show the potential at $0 \leq y \leq 0.3 $, i.e.  beyond the bifurcation point.  It is clear here that  close to $y=0$ we can see only the maximum of the potential 
at $x=0$ and an extremely unstable de Sitter space when coupled to gravity. 

\begin{figure}[h!]
\centering \epsfysize=8cm
\includegraphics[scale=0.75]{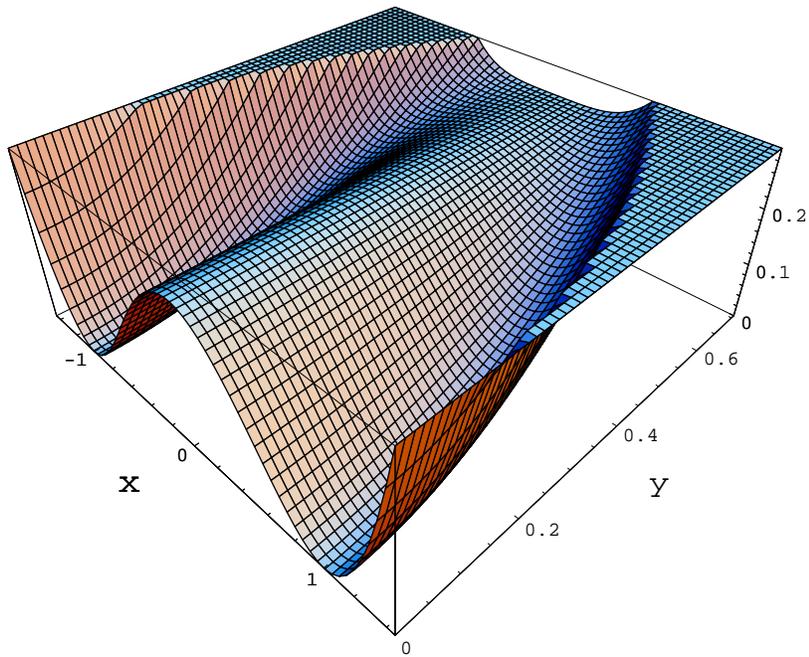}
\caption{ The potential is flat at the bifurcation point and has a maximum at $x=0$ and $y<y_b$.} \label{onefielddistrib3}
\end{figure}

\begin{figure}[h!]
\centering \epsfysize=8cm
\includegraphics[scale=0.7]{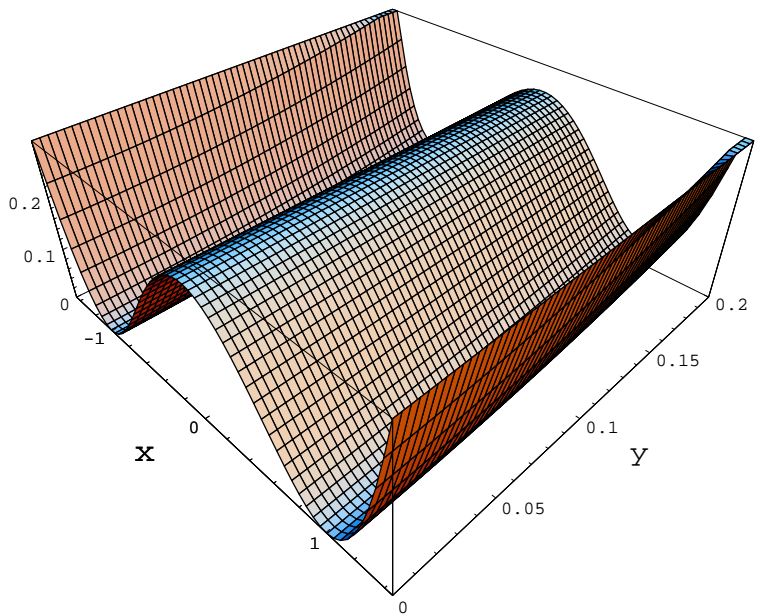}
\caption{  The potential beyond the bifurcation point shows a local maximum de Sitter stage, extremely unstable. The absolute ground state at $y=0$,~ $x^2= 1$ has 2 unbroken supersymmetries.} \label{onefielddistrib4}
\end{figure}

 \begin{figure}[h!]
\centering \epsfysize=8cm
\includegraphics[scale=1.1]{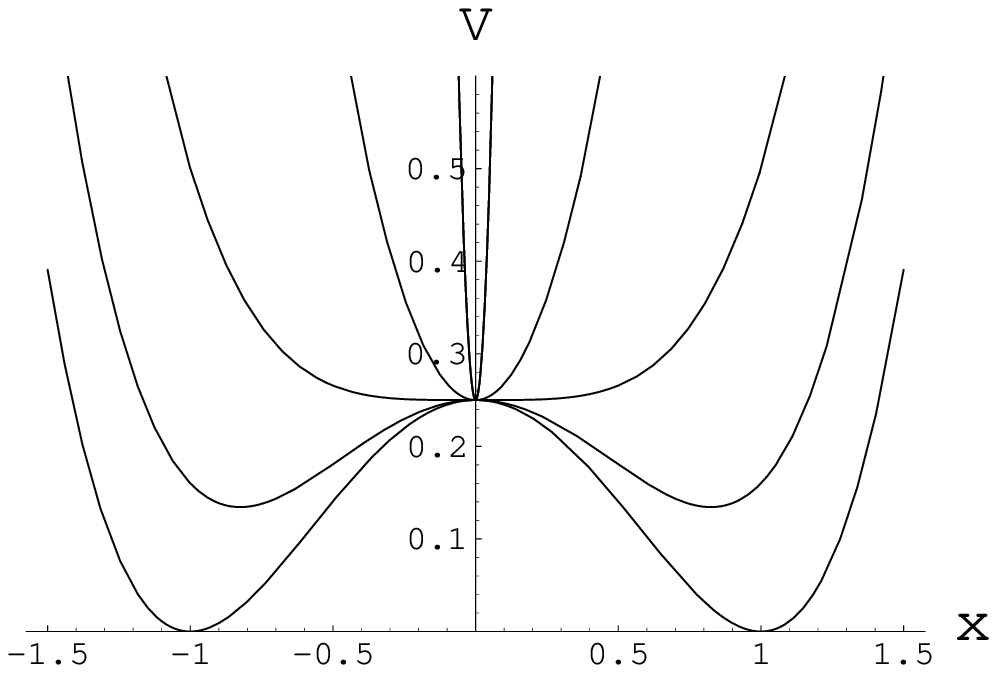}
\caption{Slices of the potential of the hybrid hypersymmetric model, from $y = 0$ to $y = 10$. The lower curve corresponds to $y= 0$. The curvature becomes positive for $y > 1/\sqrt 2$.} \label{onefielddistrib5}
\end{figure}

Finally, in Fig. 5 we have plotted the slices of the potentials at $y=10,  2, y_b, 0.3, 0$. The first 2 curves at $x=0$ have large positive curvature and positive potential, corresponding to the local de Sitter valley.
At $y=y_b$ and $x=0$ the curvature is flat, the minimum is changing into the maximum,  it is negative for $y=0.3$,  $x=0$ and finally,
at $y=0$ the curve reaches the vanishing value of the potential at $x^2= 1$.


\section{Coupling gauge theories to supergravity}

\subsection{Potentials of N=2 gauged supergravities}

Coupling of N=2 supersymmetric gauge theories to supergravity was developed over the last 20 years, see for example \cite{deWit:1985px,Andrianopoli:1996vr}. One would expect that at least in principle, any N=2 supersymmetric gauge theory can be coupled to N=2 supergravity. The opposite route, from gauged N=2 supergravity to a rigid N=2 gauge theories was developed in \cite{Andrianopoli:1996vr}, where the proper limit of $M_{P}\rightarrow \infty$ was presented. In the recent investigations of superconformal symmetry, supergravity and cosmology in \cite{Kallosh:2000ve} it was found that such limit is greatly simplified in the context of the underlying superconformal theory.

More recently there was a renewal of the interest to gauged supergravities and their potentials in a search of domain wall solutions. The least understood part of N=2 gauged supergravities is a quaternionic part. Some new studies of the hypermultiplet structures of N=2 supergravities in \cite{deWit:2001dj,D'Auria:2001kv,Ceresole:2001wi,deWit:2001bk,Alekseevsky:2001if} have revealed interesting features of the theory, which were not known before. Here we will give a short summary on this to the extent to which it is relevant to our search of de Sitter solutions in gauged N=2 supergravities.

Gauged N=2 supergravity with general interactions of vector multiplets and hypermultiplets is defined by some special K\"{a}hler and quaternionic geometries with some gauge isometries. The procedure of gauging involves the following major steps: i) the derivatives on the fields become covariant derivatives with the gauge coupling constant $g$ (different for each gauge group). These derivatives are defined in terms of Killing vectors related to isometries of the special K\"{a}hler and quaternionic manifolds, ii) for preservation the supersymmetry the action of ungauged supergravity after gauging has to be supplemented by two terms, the fermion `mass matrix' and a potential. At $g=0$ all these terms disappear and ungauged supergravity is recovered. The  potential can be presented as follows \cite{Andrianopoli:1996vr}:
\be
V_{\rm supergr}^{N=2} = - 3 {\bar L}^\Lambda  L^\Sigma P_\Lambda^r P_\Sigma^r + g^2 (g^i{}_{j} k^{\Lambda i} k^{* j}_\Sigma +4 h_{uv}k^u_\Lambda k^v_\Sigma){\bar L}^\Lambda  L^\Sigma + g_i{}^j f^{\Lambda i}  f^{*\Sigma}_j \sum _{r=1}^{r=3} P_\Lambda^r P_\Sigma^r \ .
\label{N2sugra}\ee
Here $L^\Lambda(z, \bar z)$ are the covariantly holomorphic sections of the special geometry, $f^{\Lambda i}= \nabla^i L^\Lambda$ and $P^r_\Lambda(z, \bar z, q)$ are the triplets of the prepotentials for each gauge group. There are $n$ vector multiplets and a graviphoton from the supergravity multiplet, $\Lambda = 0,1, \dots , n $. The potential is not positive definite, the first term, which was absent in the rigid case is negative definite. A useful form of the potential is given \cite{D'Auria:2001kv} in terms of a  gravitino mass matrix of N=2 theory $S_A{}^B$ related to prepotential 
\be
L_{\rm gravitino}^{\rm mass}= 2 g  \bar \psi^A_\mu  S_{AB}\gamma^{\mu\nu} \psi_{\nu}^B \ , \qquad S_A{}^B \equiv  i P^r_\Lambda \sigma_A{}^B L^\Lambda  \ .
\ee
The potential can be rewritten as follows:

\be
V_{\rm supergr}^{N=2} = g^2[ - 6 S_{AB} (S^{AB})^* + 2  g_{i}{}^j \nabla^{i} S_{AB} \nabla_{*j} (S^{AB})^*  + 4 \nabla_u S_{AB} \nabla^u S_{AB}^* +  g^{i}{}_j k_i k^{*j}]\ .
\label{sugra2}\ee
where $k_i= k_{i\Lambda} L^\Lambda$ and $k^{*j}= k^{*j}_\Sigma \bar L^\Sigma$. The negative definite contribution to the potential comes from the first term, the square of the gravitino mass term. The rest is positive definite, it includes the square of the derivatives over scalars of the special geometry as well as hypers and a terms with Killing vectors. 
This formula is quite intuitive and more easily associated with the formula for N=1 potential. In N=1  there is only one  gravitino and therefore the mass matrix has just one element,  ${W\over M_P^2} e^{K\over 2M_P^2} $.
The negative contribution to the potential  also comes from the gravitino mass term, see eq. (\ref{Vtotal}). In the proper limit when $M_P \rightarrow \infty$ the gravitino mass vanishes and the potential of the rigid N=2 theory is positive definite.

Some examples of N=2 potentials will be studied in a separate publication \cite{Kallosh:2001gr}. The problem which we would like to address here is whether  it is possible to couple hybrid hypersymmetric gauge theory to N=2 supergravity and preserve some nice features of the potential, discussed before.

\subsection{FI terms and topological obstruction to couple hybrid hypersymmetry model to N=2 supergravity} 

The cosmologically interesting feature of the hybrid hypersymmetry model is due to the existence of the constant values of a component of a prepotential triplet in the $U(1)$ gauge theory.

It is a well known fact that in N=1 supersymmetry only in the abelian gauge group sector the theory can have  constant FI D-terms. Adding a term linear in auxiliary field $D$ to the action would break gauge symmetry and supersymmetry. It will be useful for our purpose to explain this using a particular property of K\"{a}hler manifolds. The Killing vectors $k_{\Lambda i}$ on a  K\"{a}hler manifolds are the derivatives of the Killing prepotentials $P_{\Lambda}$:
\be
k_{\Lambda i}= i g_i{}^j\partial_{*j}P_{\Lambda} \ .
\ee
The prepotentials  satisfy a Poisson bracket relation:
\be
\{ P_\Lambda \ , P_\Sigma \}= f_{\Lambda \Sigma}^{\Gamma} P_{\Gamma} \ .
\label{Poisson}\ee
In the non-Abelian case this presents an obstruction to a possibility to add constant terms to the prepotentials.  It also sets a precedent for us that a constant positive contribution to the   potentials may be difficult to have in agreement with supersymmetries and gauge symmetries. In abelian case there is no problem since the eq. (\ref{Poisson}) has a vanishing right hand side,
\be
\{ P_\Lambda \ , P_\Sigma \}= 0 \ .
\ee
and shift to a constant in any direction $\lambda$ does not contradict
this relation.

Without the FI term, the SSF model of   \cite{Salam:1975wa,Fayet:1976yi} can be
coupled to supergravity as it was done in \cite{deWit:1980gt,deWit:1985px}.  
However, in presence of FI term there is so far no coupling to N=2 supergravity of this model available, to the best
of our understanding.  Moreover, we will present below the argument, that indicates that no such coupling may be possible in presence of constant $|P^3|=\xi$ terms.

To explain the argument we have to remind that the hypermultiplets of a rigid N=2 supersymmetry are associated with  hyper-K\"{a}hler manifolds, whereas in N=2 supergravity they are associated  with quaternionic manifolds \cite{Bagger:1983tt}. A hyper-K\"{a}hler manifold has a flat $SU(2)$ curvature, and a  quaternionic manifold  has a non-vanishing $SU(2)$ curvature ${\cal R}^r_{uv}$. The relation between the Killing vectors of quaternionic manifolds $k^v_{\Lambda}$ and the triplets of the  prepotentials $P^r_\Lambda$ known for a long time, see \cite{Galicki:1987ja},  is of the following nature:
\be
{\cal R}^r_{uv} k^v_{\Lambda}= D_u P^r_\Lambda\ , \qquad k_\Lambda ^u= -{4\over 3} {\cal R}^{r\,uv}D_v P^r_\Lambda \ .
\ee
Here $D_u$ is an $SU(2)$ covariant derivative over the quaternions
\be
D_u P^r_\Lambda\equiv \partial_u P^r_\Lambda + 2 \varepsilon ^{rst} \omega_u^s P^t_\Lambda \ ,
\ee
and the $SU(2)$ curvature is ${\cal R}^r= d\omega^r -  \varepsilon ^{rst}  \omega^s \omega^t$. These prepotentials have recently been studied in a superconformal
context in \cite{deWit:1998zg,deWit:2001bk}. The new relation which was found in \cite{Ceresole:2001wi} states that in local N=2 supersymmetry, i.e. in N=2 supergravity, the {\it prepotentials are defined uniquely from the Killing vectors:}
\be
P^r_\Lambda= {1\over 4 n_{H}}D_u k_{\Lambda v} {\cal R}^{r uv} \ ,
\label{obstruction}\ee
where $n_{H}$ is the number of hypermultiplets. On the basis of this relation it has been concluded in \cite{Ceresole:2001wi} that any covariantly constant shift $P^{(0)r}_\Lambda$ is excluded since 
the integrability condition, $\varepsilon^{rst} {\cal R}^{s uv}P^{(0)t_\Lambda}=0$ implies that  $P^{(0)r}_\Lambda=0$. This implies the absence of the FI terms in all cases of N=2 supergravity with hypermultiplets. The exceptional cases when FI terms are possible, include theories with vector multiplets without hypermultiplets or cases of rigid supersymmetry when the curvature of the hyper-K\"{a}hler manifold vanishes. In both cases ${\cal R}^r_{uv}=0
$ 
and FI terms are possible in agreement with supersymmetry and gauge symmetry. The confirmation of this analysis comes from the recently discovered \cite{D'Auria:2001kv} harmonicity property of the quaternionic prepotential
\be
D^u D_u P^r_\Lambda= 2 n_{H} P^r_\Lambda \ .
\ee
This equation can be used to derive the equation (\ref{obstruction}).
Thus in view of the properties of the quaternionic geometries explained above it looks plausible that the consistent coupling of hybrid hypersymmetric model with constant $\xi$ to N=2 supergravity is not possible. However, more investigations in this directions would be required to make a conclusive statement \footnote{One possibility to overcome this topological obstruction has been suggested to us by A. Van Proeyen, private communication. 
Suppose that there is some gauging
in N=2 sugra which  has the prepotential $P^r$ that is not constant.
Still, in the limit $M_P \rightarrow \infty$  it might tend to  a constant value. In such case one may hope to find the embedding of the hybrid hypersymmetric model into N=2 supergravity with de Sitter
or near de Sitter vacuum.}. The $SU(2,2|2)$ superconformal part of the model  can be consistently coupled to N=2 supergravity, only the constant $\xi$-terms violating the superconformal symmetry seem to cause the problem \footnote{We understand that the problem of coupling of SSF model with FI terms to N=2 supergravity is now under investigation by B. de Wit and S. Vandoren, private communication.}.

\subsection{Coupling of hybrid hypersymmetry model to N=1 supergravity}

Coupling of hybrid hypersymmetric model with rigid N=2 supersymmetry to N=1 supergravity is simple. We may ignore the fact that the rigid limit has double supersymmetry and proceed as if only one supersymmetry is available and make it local. This procedure is of course not unique. The simplest one which we will also follow here was proposed in the context of D-term inflation in \cite{Binetruy:1996xj}. One can choose the minimal K\"{a}hler manifold for all 3 chiral superfields, two from the hypermultiplet, $\Phi_1, \Phi_2$ and  the one from the vector multiplet $ \Phi_3$. 
In the K\"{a}hler geometry all 3 chiral multiplets now enter in a symmetric way. The D-terms will reflect that $\Phi_1$ and $\Phi_2$ have opposite charges and $\Phi_3$ is neutral.
Also the kinetic term function $f$ of the vector multiplet may be taken minimal, i.e. field independent delta-function. Thus our N=1 supergravity has the 
K\"{a}hler potential, the superpotential and the potential given by
\be 
K=\sum_{a=1}^{a=3} |\Phi_a|^2 \ , \qquad W= {g\over 6 \sqrt 2} \varepsilon^{abc}  \Phi_a \Phi_b \Phi_c \ .
\ee 
\bea
V &=& {g^2\over 8} e^{|\Phi_a|^2\over M_P^2} \left( \left[
|\varepsilon^{abc} \Phi_b \Phi_c| \left (1 + {|\Phi_a|^2\over M_P^2}\right)^2 \right]^2-{1\over 3 M_P^2}|\varepsilon^{abc}  \Phi_a \Phi_b \Phi_c|^2 \right) \nonumber\\
&+& {g^2\over 8}(|\Phi_1|^2 -|\Phi_2|^2+ {2\xi\over g} )^2] \ .
\eea
Note the negative contribution due the term with the square of gravitino mass,  $ -3 e^{K\over M_P^2}|W|^2$.
Let us consider the critical points of this potential.

1. {\it Non-supersymmetric de Sitter valley} at some  undefined but restricted constant value of the scalar $\Phi_3$, all hypers  vanish.
\be
|\Phi_3|= (|\Phi_3|)_0\ , \qquad    \Phi_1=  \Phi_2=0\ , \qquad W=0 \ , \qquad {\cal D}_a W=0  \ , 
\ee
and 
\be 
 D=- \xi \ , \qquad V= { \xi^2 \over 2} \ .
\ee
This is a local minimum of the potential,  all supersymmetries are broken. Einstein equations have  {\it de Sitter solution}. The possibility to have constant D-terms in N=1 supergravity was known for a long time. More recently a superconformal origin of FI terms was clarified in \cite{Kallosh:2000ve}. It was found there that the D-terms appear via the gauge transformation of the conformon multiplet, which plays an important role in the superconformal theory underlying supergravity.

2. {\it Supersymmetric absolute global minimum}
\be
(\Phi_3)_{\rm susy}= (\Phi_1)_{\rm susy}= 0 \ ,  \qquad W_{\rm susy}=0 \ , \qquad ({\cal D}_a W)_{\rm susy}=0  \ ,
\ee
and 
\be
D_{\rm susy}= {g\over 2}|\Phi_2|^2_{\rm susy}- \xi =0\ , \qquad V_{\rm susy}= 0\ .
\ee

Thus the absolute minimum ground state has  N=1 local supersymmetry unbroken and a vanishing potential.  $U(1)$ gauge symmetry is spontaneously broken.  Einstein equations at this vacuum have a  solution with {\it vanishing cosmological constant}.

\section{P-term inflation and acceleration of the universe}

We will present here some  short remarks about the applications in
cosmology of N=2 supersymmetry  with P-term inflation/acceleration,
where $P^r$ is the triplet of the Killing prepotentials of N=2 gauge
theory.

From our results we may conclude that at the tree level of N=2
supersymmetry of our model we have a de Sitter solution for $\Phi_2 = 0$
and for all sufficiently large values of the inflaton field $\Phi_3$.
However for the cosmological solution we need to provide a slow roll
regime so that inflation takes place.
This issue has been analysed for D-term inflation in
\cite{Binetruy:1996xj}, and
for P-term inflation the situation is very similar.
It turns out that
one can lift the flat direction of the inflaton field   due to the first
loop corrections in gauge theory.  The tree level  splitting of the
masses in supermultiplets in de Sitter vacuum leads to the effective
1-loop potential for large inflaton field $\Phi_3$:
\be
V_{1-\rm loop}= {\xi^2\over 2}\Bigl[1 + {g^2\over 8\pi^2} \ln {
|\Phi_3^2|\over {|\Phi_3^2|_c} }  +...
\Bigr] .
\ee
This term is important because it leads to the motion of the field
$\Phi_3$ towards the bifurcation point and the end of inflation. It is
interesting to note that for the N= 2 P-term inflation, which
corresponds to the N= 1 D-term inflation in the case $\lambda=\sqrt 2 \,
g$  when this theory can be embedded into N=2 supersymmetric model, {\it
all non-gravitational higher loop corrections
are finite} \cite{Grisaru:1982zh}.

As we already pointed out, even though our model is different from the
F-term inflation,  the effective potential of the fields $\Phi_2$ and
$\Phi_3$ shown in Fig. 1 of our paper looks the same as in the simplest
version of F-term inflation \cite{Dvali:1994ms}.  This theory has an
interesting property. As one can see  from Fig. 1, the potential near
the bifurcation point has a very complicated shape. Thus one could
expect that after the field $\Phi_3$ reaches the bifurcation point, both
fields $\Phi_2$ and $\Phi_3$ will fall towards the minimum of the
effective potential along a very complicated trajectory. This is indeed
the case for the general D-term inflation models. Surprisingly enough,
in the F-term model \cite{Dvali:1994ms} the fields roll from the
bifurcation point to the minimum of the effective potential  along a
straight line \cite{Bastero-Gil:1999fz}. This simple behavior of the
fields occurs in  D-term inflation only for the  particular case
$\lambda=\sqrt 2 \, g$  corresponding to N= 2 P-term inflation:  After
inflation the fields $\Phi_3$ and $\Phi_2$ simultaneously oscillate
along the straight line $ \Phi_2 + \sqrt 2 \Phi_3= \sqrt \xi$.

To study inflation in this theory one should use the Friedmann equation
\begin{equation}\label{infl}
H^2 = \left({\dot a \over a}\right)^2 = V/3 \approx {\xi^2\over 6},
\end{equation}
where $a(t)$ is a scale factor of the universe. Thus one has $H =
\xi/\sqrt 6$.
This leads to inflation
\begin{equation}\label{infl1}
a(t) = a(0)~ \exp{ {{\xi\  t \over\sqrt 6} }}~ .
\end{equation}
For simplicity of notation, in the description of this stage we will
write $\phi$ instead of $|\Phi_3|$.
During the slow-roll regime the field $\phi = |\Phi_3|$ obeys equation
$3H\dot\phi = -V'(\phi)$ \cite{book},
which gives
\begin{equation}\label{infl2}
\phi^2(t) = \phi^2(0) - {g^2\xi ~t\over 2\sqrt 6 \pi^2} \ .
\end{equation}

Suppose inflation ends soon after the field $\phi$ becomes smaller than
$\phi_c = |\Phi_3|_c = \sqrt{\xi/g}$. This is a generic property of
almost all versions of hybrid inflation \cite{Linde:1991km}. Then using
equations (\ref{infl1}), (\ref{infl2}) one can find the value of the
field $\phi_N$ such that the universe inflates $e^N$ times when the
field rolls from $\phi_N$ until it reaches the bifurcation point:
\begin{equation}\label{infl3}
\phi_N^2  = {\phi_c}^2  + {g^2 N\over 2 \pi^2}.
\end{equation}
Density perturbations on the scale of the  present cosmological horizon
have been produced at $\phi \sim \phi_{N}$ with $N \sim 60$, and their
amplitude is proportional to ${V^{3/2}\over V'  }$ at that time
\cite{book}.
One can find parameters of our model using COBE normalization for
inflationary perturbations of metric on the horizon scale
\cite{Linde:1997sj,Lyth:1999xn}:
\begin{equation}\label{smallg}
 {V^{3/2}\over V'  }= {2\sqrt 2\pi^2 \xi\over  g} \phi_N \sim
5.3 \times 10^{-4},
\end{equation}
where $N \sim 60$.
If ${\phi_c}^2 = {\xi\over g} \ll {g^2 N\over 2 \pi^2}$, one has $
\phi_N   =  {g \sqrt{N}\over \sqrt 2 \pi },$
and
\begin{equation}\label{COBE1}
 {V^{3/2}\over V'  }= {2\pi\over  g} {\xi \sqrt N} \sim
5.3 \times 10^{-4}.
\end{equation}
For $N \sim 60$ this implies that
\begin{equation}\label{COBE2}
{\xi\over g}  \approx 1.1 \times 10^{-5}  \ .
\end{equation}
One can represent this result in terms of the amplitude of spontaneous
symmetry breaking of gauge symmetry:
\begin{equation}\label{COBE2}
|\Phi_2| = \sqrt 2|\Phi_3|_c = \sqrt {2\xi\over g} \approx 4.7 \times
10^{-3} M_p \approx \ 1.1\times 10^{16}~ {\rm GeV}.
\end{equation}
This is very similar to the GUT scale.

It is worth mentioning that in addition to inflationary perturbations
there may appear
perturbations of metric created by cosmic strings that are formed after
the end of inflation in this scenario
\cite{Kofman:1987wm,Linde:1991km}. An investigation of this issue in
the  simplest versions of F-term and D-term inflation have shown that
cosmic strings may lead to perturbations of metric proportional to
$O({\xi/g})$. These perturbations can be of the same order of magnitude
as inflationary perturbations  or even few times greater
\cite{Jeannerot:1997is,Linde:1997sj,Lyth:1999xn}. If one wants to avoid the cosmic
string contribution, one may  suppress production of strings and other
topological defects  by making certain modifications to the model, see
e.g. \cite{Lazarides:2000ck} and references therein. Alternatively, one
may consider the models with  very small  $g^2$.

Indeed,  the  results described above have been obtained for $ \phi_c^2
=
{\xi\over g} \ll    {g^2 N \over  2 \pi^2 } \sim 3g^2$, i.e. for $g^2
\gg 3 \times 10^{-6}$. This is a natural assumption, but one should note
that $g^2$ in our model is not related to the gauge coupling constant in
GUTs, so it can be very small.
In the regime $g^2 < 3\times  10^{-6}$ one has $\phi_N \approx \phi_c=
\sqrt{\xi\over g}$, so that
\begin{equation}\label{smallg}
 {V^{3/2}\over V'  }= {2\sqrt 2\pi^2\over  g} \left({\xi \over
g}\right)^{3/2} \sim
5.3 \times 10^{-4}.
\end{equation}
This yields
\begin{equation}\label{smallg}
{\xi \over g} \sim 0.7\times 10^{-4}~ g^{2/3}.
\end{equation}
For $g^2 \ll 3\times 10^{-6}$ one has ${\xi \over g} \ll 10^{-5}$. This
suppresses the cosmic string contribution to density perturbations while
keeping the inflationary contribution intact.

In general, the hybrid hypersymmetry model  can be used not only for
inflation in the very  early universe but also to explain the observed
acceleration of the universe at the present epoch. Depending on the
parameters of the theory and on initial conditions for the inflaton
field $\Phi_3$, the universe may spend a very long time
in   de Sitter valley $\Phi_2 = 0$, $|\Phi_3| > \sqrt{\xi\over g}$. The
bifurcation point may be approached very slowly but  still within a
finite time. When it is reached, as Figures 1-5 show, the waterfall
stage will bring the universe to an absolute ground state with the
vanishing cosmological constant and unbroken hypersymmetry. Recently it
was argued that in a class of theories with a stable supersymmetric
vacuum, a system cannot relax into a zero-energy supersymmetric vacuum
while accelerating if the evolution is dominated by a single scalar
field with a stable potential \cite{Hellerman:2001yi}. This was
considered as a challenge for string theory as presently defined. In our
model this problem does not appear: acceleration of the universe is not
eternal, and the universe enters the state with unbroken supersymmetry
due to the combined motion of two scalar fields.

One of the  features of hybrid hypersymmetry model is that
it requires at least 3 complex scalars: 1 complex scalar  in a vector
multiplet and a quaternion in a hypermultiplet. We were able to obtain
de Sitter solution in N=2 theory  only in presence of  FI P-terms, which
leads to hybrid inflation  \cite{Linde:1991km}.
Thus  hybrid inflation  may appear to be the only mechanism for
inflation/acceleartion  consistent with N=2 supersymmetry.

We will show in \cite{Kallosh:2001gr} that all potentials available in the
literature in N=2,4,8 gauged supergravities have de Sitter solutions
which are either  maxima or  saddle points, in all cases highly
unstable. However, most of the scalars in these gauged supergravities
were truncated. Note that if one would take $|\Phi_3| < \sqrt{\xi\over
g} $  in our N=2 model, one would also obtain only unstable solutions,
as shown in Fig. 3,4.
It is plausible that  keeping more scalars (there are 35 complex scalars
in N=8 theory) one can have a better chance to recover some hybrid-type
potential with  a long lasting de Sitter stage suitable for inflation
and/or acceleration.

Another interesting possibility may be that the hybrid hypersymmetry
gauge model with potential shown in Figures 1-5,  may be directly
related to string theory on $adS_5\times M^5$ space or to some stringy brane construction of the relevant gauge theory.
This suggests that a further analysis of this and other theories with
$N\geq 2$ supersymmetries   may lead to interesting new links between
M/string theory and cosmology.

\

I am grateful to O. Aharony, C. Herdeiro, S. Hirano, C. Hull, S. Kachru,
L. Kofman, A. Linde, A. Maroto, S. Prokushkin, M. Shmakova, E.
Silverstein, S. Susskind,  S. Vandoren, A. Van Proeyen, P. West, B. de
Wit and E. Witten for stimulating discussions. This work is supported by
NSF grant PHY-9870115.

\end{document}